\documentclass[a4paper,11pt]{article}
\usepackage{pos}

\usepackage{caption}
\usepackage{subcaption}
\usepackage{slashed}

\title{On the $ZH$ amplitudes at two loop in QCD}

\author{Taushif Ahmed}

\affiliation{Dipartimento di Fisica and Arnold-Regge Center, Universit\`a di Torino, 
\\ and INFN, Sezione di Torino\\ Via Pietro Giuria 1, I-10125 Torino, Italy}

\emailAdd{taushif.ahmed@unito.it}

\abstract{Through this article, we show that the amplitudes for the $b\bar{b} \rightarrow ZH$ in the presence of bottom-Higgs Yukawa coupling can be constructed, up to a few anomalous diagrams, solely from a set of vector form factors of properly grouped classes of diagrams. Thereby, we can completely bypass the subtle issues involving the chiral quantity $\gamma_5$. In the second part, we demonstrate that while computing the $q\bar{q} \rightarrow ZH$ amplitudes in Higgs effective field theory employing a non-anticommuting $\gamma_5$ in dimensional regularization, we need to introduce a four-point effective composite operator to the renormalized Lagrangian in order to retain the physical properties of the amplitude.}

\FullConference{%
  *** The European Physical Society Conference on High Energy Physics (EPS-HEP2021), ***\\
  *** 26-30 July 2021 ***\\
  *** Online conference, jointly organized by Universität Hamburg and the research center DESY ***
}

\tableofcontents

\begin{document}
\maketitle

\section{Introduction}
\label{sec:intro}
VH events play an important role in the exploration of the Higgs physics at the Large Hadron Collider (LHC). Understanding the fundamentals of the electroweak symmetry breaking from direct experiment is of prime importance for which a detailed exploration is underway. The recent observation~\cite{Aaboud:2018zhk,Sirunyan:2018kst} of the Higgs boson decaying to a pair of bottom quarks gets its primary contribution from the VH channel. Due to its high phenomenological importance, there has been a significant number of computations available in the literature aiming to make the theoretical predictions more precise~\cite{Brein:2003wg,Brein:2011vx,Ahmed:2014cla,Ferrera:2014lca,Li:2014bfa,Catani:2014uta,Kumar:2014uwa,Campbell:2016jau,Ferrera:2017zex,Ahmed:2019udm}. 

In this article, we consider the two-loop QCD corrections for the production of the scalar Higgs ($H$) in association with the $Z$ boson through bottom quark annihilation in presence of bottom-Higgs non-zero Yukawa coupling. We show~\cite{Ahmed:2020kme} that the non-anomalous part of the axial amplitude can be obtained by completely bypassing an explicit computation of the axial part. This means we can completely avoid encountering the subtle issues involving the chiral quantity $\gamma_5$.

In the second part of the article, we show that while computing the two-loop QCD corrections to the production of $HZ$ in Higgs effective field theory (HEFT), we are bound to encounter a striking phenomenon. In particular, we find that the amplitude computed using non-anticommuting $\gamma_5$ in dimensional regularization fails to exhibit the expected chiral Ward identity and universal infrared behaviour. To restore these properties, we need to introduce an amendment involving four-point effective operator to the renormalized Lagrangian. On the other hand, if we apply an anticommuting $\gamma_5$, we do find the amplitudes fulfil all the expected properties.

\section{Axial vector form factors from vector counterparts for $b\bar b \rightarrow ZH$ amplitudes}
We consider the production of massive $Z$ and scalar Higgs ($H$) bosons through bottom ($b$) quark annihilation:
\begin{align}
\label{eq:process}
    b(p_1) + \bar{b}(p_2) \to Z(q_1) + H(q_2)\,.
\end{align}
The quantities within the parentheses denote the corresponding four-momentum satisfying $p_1^2=p_2^2=0,\,q_1^2=m_Z^2,\,q_2^2=m_H^2$, where $m_Z$ and $m_H$ are respectively the masses of $Z$ and $H$ bosons. We define the Mandelstam variables as $s\equiv (p_1+p_2)^2, t\equiv (p_1-q_1)^2, u\equiv (p_2-q_1)^2$ with $s+t+u=q_1^2+q_2^2=m_Z^2+m_H^2$. We keep a non-zero $b\bar{b}H$ Yukawa coupling $\lambda_b$, otherwise $b$ quark is treated as
\begin{figure}[htbp]
\begin{center}
\includegraphics[scale=0.24]{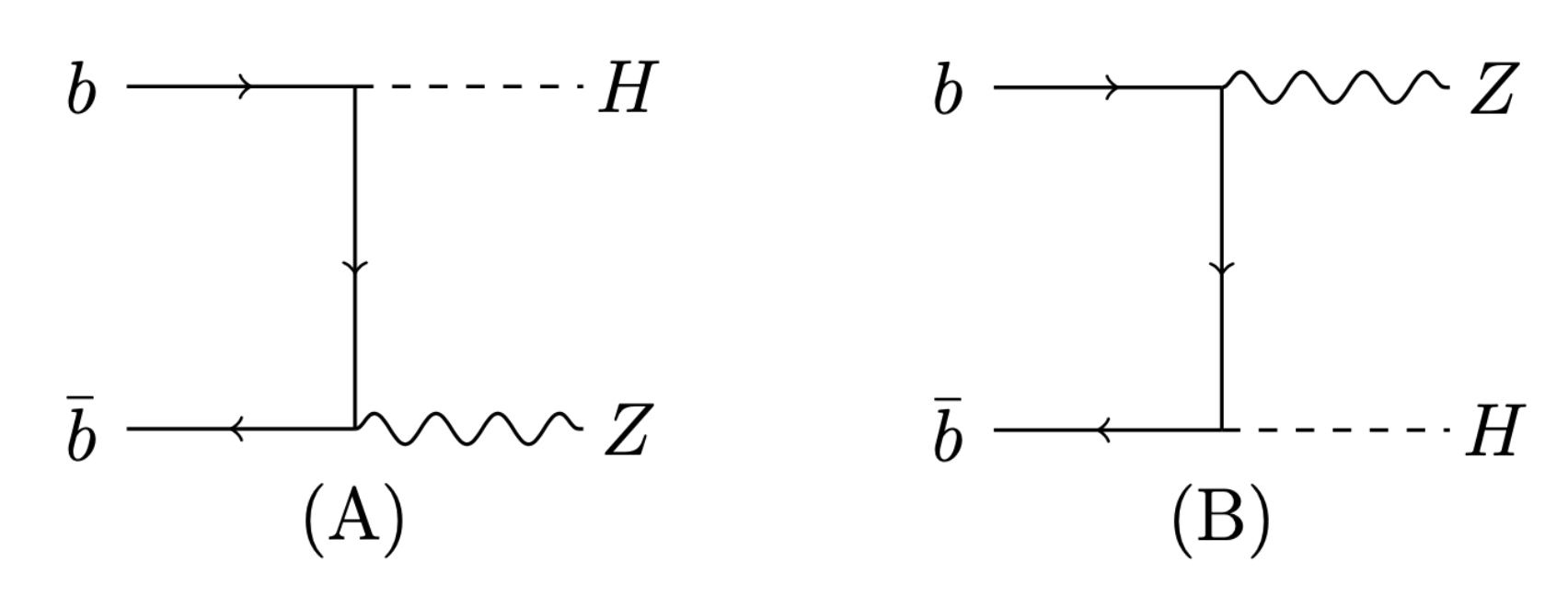}
\caption{Leading order Feynman diagrams that involve the bottom-Higgs Yukawa coupling.}
\label{dia:tree}
\end{center}
\end{figure}
massless. The top quark is considered as infinitely heavy and hence we work in $n_f=5$ massless flavor limit. In this article, we compute the two-loop scattering amplitude in QCD for the non-Drell-Yan type diagrams, as shown in figure~\ref{dia:tree}. We parameterize the scattering amplitude as
\begin{align} 
\label{eq:bbZHamplitude}
    \mathcal{M} &= \lambda_b\, \bar{v}(p_2) \, \mathbf{\Gamma}^{\mu} \, u(p_1) \, \varepsilon^{*}_{\mu}(q_1) \nonumber\\
    &=\bar{v}(p_2) \left( \lambda_b~ g_{V,b}   \, \mathbf{\Gamma}^{\mu}_{vec}  
    \,+\,\lambda_b~ g_{A,b} \, \mathbf{\Gamma}^{\mu}_{axi} \right) u(p_1) \, \varepsilon^{*}_{\mu}(q_1)\nonumber\\
    &\equiv  \lambda_b~ g_{V,b} {\cal M}_{vec}+ \lambda_b~ g_{A,b} {\cal M}_{axi}\,.
\end{align}
The symbol $\mathbf{\Gamma}^{\mu} \equiv  g_{V,b}\mathbf{\Gamma}^{\mu}_{vec} +  g_{A,b}\mathbf{\Gamma}^{\mu}_{axi}$ represents 
a matrix in the spinor space with the vector and axial vector couplings $g_{V,b}$ and $g_{A,b}$, respectively. $\varepsilon^{*}_{\mu}$ is the polarization vector of the $Z$ boson. 
Our \textit{goal} is to demonstrate an elegant formalism~\cite{Ahmed:2020kme} that allows us to get the non-anomalous axial contributions by bypassing the explicit computation. In particular, we show how to get this contribution from the vector part of the amplitude.

\begin{figure}[htbp]
\begin{center}
\includegraphics[scale=0.45]{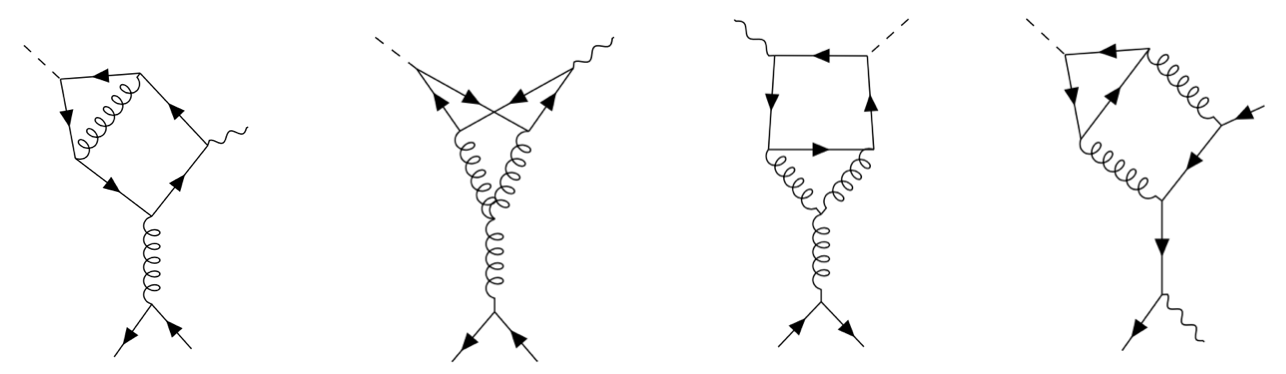}
\caption{Sample diagrams where the Higgs or Higgs+$Z$ bosons couple to a quark loop.}
\label{dia:H-closed-fermion}
\end{center}
\end{figure}
We employ an anticommuting $\gamma_5^{\rm AC}$ in $D=4-2\epsilon$ dimensions. Note that the set of diagrams, as shown in figure~\ref{dia:H-closed-fermion}, where either Higgs or Higgs+$Z$ bosons get radiated from closed fermionic loop identically vanish owing to the trace over an odd number of $\gamma$ matrices. Therefore, at two-loop we get non-zero contributions only from those diagrams where $Z$ boson couples to the open $b$ quark line and consequently the $\gamma_5^{\rm AC}$ in $\mathbf{\Gamma}^{\mu}_{axi}$ can be anticommuted next to an external $b$ quark spinor. This non-zero contribution is categorised as class-ZH and class-HZ corresponding to the QCD corrections to the tree level diagrams (A) and (B), respectively in figure~\ref{dia:tree}. The reason we perform this separation is because of the presence of a chirality flipping Yukawa interaction on $b$ quark line a relative negative sign gets generated between these two class of contributions when we push the $\gamma_5^{\rm AC}$ next to the same external $b$ quark spinor. This is also the reason why the vector and non-anomalous axial form factors are not identical to each other.
Let us focus solely on the vector part of the amplitude and decompose it as 
\begin{align} 
\label{eq:bbZHamplitude_VECS}
    \mathcal{M}_{vec} &= 
    \bar{v}(p_2) \left( \mathbf{\Gamma}^{\mu}_{ZH} + \mathbf{\Gamma}^{\mu}_{HZ} \right) \, u(p_1) \, \varepsilon^{*}_{\mu}(q_1)  \nonumber\\
   \mathbf{\Gamma}^{\mu}_{X}  &= 
    F_{1,X}  p_1^{\mu} 
    + F_{2,X} p_2^{\mu} + F_{3,X} q_1^{\mu}
    + F_{4,X} \gamma^{\mu} \slashed{q}_1 \,, \quad X=ZH, HZ\,,
\end{align}
where we denote the scalar form factors by $F_{i,X}$. We can calculate the set of projectors~\cite{Ahmed:2019udm,Chen:2019wyb} for this aforementioned tensor decomposition. In the absence of axial coupling, the above four tensorial structures are linearly independent and complete in $D$ dimensions for $\mathcal{M}_{vec}$ irrespective of the QCD loop order.

By applying the projectors we get the scalar form factors without encountering any subtlety related to the $\gamma_5$. The renormalisation of the bare form factors is also straightforward as dictated by the vector part of the amplitude. The complete vector form factors $F_{i,vec}$ are obtained through
\begin{align} 
\label{eq:bbZHamplitude_vecs}
F_{i,vec} =  F_{i,ZH} + F_{i,HZ} \quad \text{for} \; i = 1,2,3,4 \,.
\end{align}

Now let us come back to the axial part. We can decompose it into non-anomalous or non-singlet and anomalous or singlet contributions
\begin{equation}
\label{eq:axisns}
   {\cal M}_{axi} = {\cal M}_{axi(ns)} + {\cal M}_{axi(s)} \, .
\end{equation}
Similar to the vector part, we perform a Lorentz covariant decomposition of the non-singlet part of the amplitude by applying $\gamma_5^{\rm AC}$ in $D$ dimensions which reads as
\begin{align}
\label{eq:complete-axi-FF}
\mathcal{M}_{axi(ns)}^{\mu} &= \bar{v}(p_2) \left(
    F_{1,axi(ns)}  p_1^{\mu}
    + F_{2,axi(ns)} p_2^{\mu} 
    + F_{3,axi(ns)} q_1^{\mu}
    + F_{4,axi(ns)} \gamma^{\mu} \slashed{q}_1 \right) \gamma_5 u(p_1) \,.
\end{align}
We find that the aforementioned scalar form factors can be obtained from the vector parts through
\begin{align} 
\label{eq:bbZHamplitude_axis}
F_{i,axi(ns)} = F_{i,HZ} - F_{i,ZH}\,.
\end{align}
The reason behind the appearance of the relative negative sign is discussed in the text above \eqref{eq:bbZHamplitude_VECS}. In ref.~\cite{Ahmed:2019udm}, we explicitly computed the axial form factors by treating $\gamma_5$ appropriately. While comparing the ultraviolet renormalized finite remainders, we find a perfect  agreement. Note that the ultraviolet renormalization that is required to obtain \eqref{eq:bbZHamplitude_axis} is identical to that of the vector counterpart. The results of the  partial form factors $F_{i,ZH}$ and $F_{i,HZ}$ can be found from the ancillary files of ref.~\cite{Ahmed:2020kme}. The singlet part of the amplitude can not be restored by this method, it must be computed explicitly by applying a non-anticommuting $\gamma_5$ prescription, as demonstrated in ref.~\cite{Ahmed:2019udm}.

\section{A pitfall in $q\bar{q} \rightarrow ZH$ amplitudes in Higgs effective field theory}
\label{sec:qqzh}

\begin{figure}[h]
	\centering
	\begin{subfigure}[c]{.2\textwidth}
		\raisebox{-\height}{\includegraphics[width=0.8\textwidth,angle=90
		]{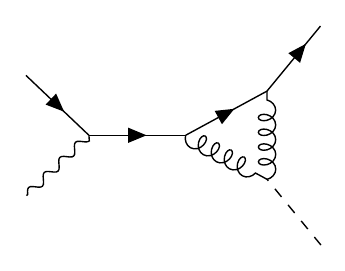}}
	\end{subfigure}
\hspace{1cm}
	\begin{subfigure}[c]{.2\textwidth}
		\raisebox{-\height}{\includegraphics[width=0.8\textwidth,angle=0]{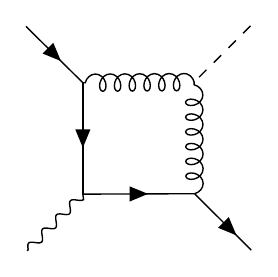}}
	\end{subfigure}
	\caption{Sample diagrams at the leading order in HEFT. The external curly and dotted lines respectively denote the $Z$ and $H$-bosons.}
	\label{fig:1L}
\end{figure}

In this section, we consider the production of $ZH$ through quark annihilation within the framework of Higgs effective field theory where the top quark loop in integrated out by treating it infinitely heavy. In figure~\ref{fig:1L}, we show some sample diagrams at the leading order.
By decomposing the amplitude $\mathcal{M}$ into a vector ($\mathcal{M}_{vec}$) and axial ($\mathcal{M}_{axi}$) parts, we can perform a Lorentz covariant decomposition in $D$ dimensions as 
\begin{align} 
\label{eq:FFdecomp-qqZH-vec}
\mathcal{M}_{vec}^\mu &= \bar{v}(p_2) \left(
    {\cal F}_{1,vec} \slashed{q_1}  p_1^{\mu} 
    + {\cal F}_{2,vec}  \slashed{q_1}  p_2^{\mu} + {\cal F}_{3,vec}   \slashed{q_1}  q_1^{\mu}
    + {\cal F}_{4,vec}  \gamma^{\mu} \right)u(p_1) \nonumber\\
\mathcal{M}_{axi}^\mu &\equiv \bar{v}(p_2) \left(
    {\cal F}_{1,axi} \slashed{q_1}  p_1^{\mu} 
    + {\cal F}_{2,axi}  \slashed{q_1}  p_2^{\mu} + {\cal F}_{3,axi}   \slashed{q_1}  q_1^{\mu}
    + {\cal F}_{4,axi}  \gamma^{\mu} \right)\gamma_5 u(p_1) \,.
\end{align}
The axial part consists of the non-singlet, $\mathcal{M}^\mu_{axi(ns)}$, and singlet, $\mathcal{M}^\mu_{axi(s)}$, contributions. While focusing on the non-singlet part and performing the calculation employing an anticommuting $\gamma_5$, we find that the finite remainders of the form factors satisfy
\begin{align}
\label{eq:Ax-equal-Vec-Eff}
    {\cal F}_{i,axi(ns)}^{(l)}={\cal F}_{i,vec}^{(l)}\,, \quad i=1,2,3,4\,, \quad l=1,2\,,
\end{align}
with 
\begin{align}
\label{eq:calF-expand}
    {\cal F}_{i,va}= \sum_{l=1}^{\infty} a^{(l+1)}_s(\mu_R) {\cal F}_{i,va}^{(l)}\,, \quad va=vec,axi(ns)\,.
\end{align}
We denote the strong coupling constant at the renormalization scale $\mu_R$ by $a_s(\mu_R)\equiv\alpha_s(\mu_R)/(4\pi)$. The identity in \eqref{eq:Ax-equal-Vec-Eff} leads to the expected chiral invariance. Our method of computation is presented in detail in refs.~\cite{Ahmed:2019udm,Ahmed:2020kme}. Instead of applying the anticommuting (AC) $\gamma_5$, if we employ a prescription of non-anticommuting (NAC) $\gamma_5$, we encounter the violation of the above identity.

\begin{figure}[htb]
\begin{center}
\includegraphics[scale=0.30]{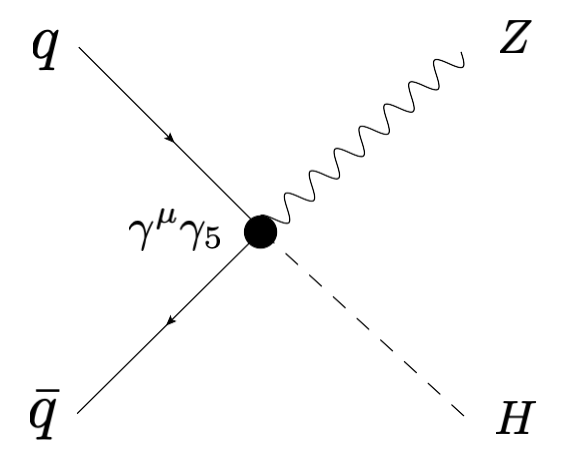}
\caption{The amendment vertex to the Lagrangian for using NAC in HEFT}
\label{dia:qqZHoperator}
\end{center}
\end{figure}

In dimensional regularization, we can define a NAC $\gamma_5$ as~\cite{tHooft:1972tcz,Breitenlohner:1977hr}
\begin{align}
\label{eq:gamma5}
	\gamma_5=-\frac{i}{4!}\varepsilon_{\mu\nu\rho\sigma}\gamma^{\mu}\gamma^{\nu}\gamma^{\rho}\gamma^{\sigma}\,,
\end{align}
where the Lorentz indices of the Levi-civita symbol $\varepsilon_{\mu\nu\rho\sigma}$ are treated in D dimensions~\cite{Larin:1991tj}. Due to this definition, we need an additional care~\cite{Larin:1991tj,Larin:1993tq,Ahmed:2021spj} while performing the ultraviolet renormalization to ensure the restoration of the Ward identity. However, surprisingly we find that the expected Ward identity gets violated while comparing the finite remainders ($\epsilon \rightarrow 0$). Although the LO vector finite remainders are identical to the corresponding quantities obtained using AC $\gamma_5$, the axial ones are not. In particular, we discover that
\begin{align}
\label{eq:FF-NAC-AC}
&\mathcal{F}_{i,axi(ns)}^{(1),{\rm NAC}}=\mathcal{F}_{i,vec}^{(1)}\,,\quad i=1,2,3,\nonumber\\
&\mathcal{F}_{4,axi(ns)}^{(1),{\rm NAC}}\neq\mathcal{F}_{4,vec}^{(1)}\,.
\end{align}
This immediately implies the violation of the Ward identity even at the LO level
\begin{align} 
\label{eq:WI_NAC5}
q_{1,\mu} \, \mathcal{M}^{\mu,{\rm NAC}}_{axi (ns)} & \neq  0\,.
\end{align}
We find that the Ward identity can be restored if we introduce an amendment term of the form
\begin{align} 
\label{eq:amendO} 
\mathcal{J}^{\mu,{\rm NAC}}
\equiv 
\mathrm{Z}_5^{h}(a_s)\, \mathrm{\mathbf{C}} \, 
\Big(\bar{v}(p_2) \,\left[\gamma^{\mu} \gamma_5\right]_{L} \, u(p_1) \Big)
\end{align}
to the renormalized Lagrangian. This amendment term can be visualised as a four-point local composite operator as shown in figure~\ref{dia:qqZHoperator}. The constant factor $\mathrm{\mathbf{C}} \equiv a_s \left(-4 C_F\right) {C_H}/{v} $ absorbs the overall $a_s^2$ of the LO amplitude. $C_F$ is the quadratic Casimir of SU($n_c$) gauge group in fundamental representation. $C_H$ and $v$ are respectively the Wilson coefficient and vacuum expectation value of the scalar Higgs.
The symbol $[]_L$ means the quantity is renormalized according to the Larin scheme~\cite{Larin:1991tj,Larin:1993tq}. We introduce an additional renormalization constant $\mathrm{Z}_5^{h}(a_s)=1+{\cal O}(a_s)$ which should be determined order by order in perturbation theory. By performing an explicit computation to the next-to-leading order, we find~\cite{Ahmed:2020kme} that the corresponding non-singlet contribution as
\begin{align} 
\label{eq:amend_Z5ns} 
\mathrm{Z}_{5, ns}^{h}(a_s) &=
1 \,+\,  a_s\, \left( 
\frac{-\beta_0}{\epsilon} 
\,+\, 
\frac{107}{18} C_A   
- 7 C_F - \frac{1}{9} n_f \right)
\,+\, \mathcal{O}(a^2_s) \, ,
\end{align}
and the singlet as
\begin{align} 
\label{eq:amend_Z5s} 
\mathrm{Z}_{5, s}^{h}(a_s) &=
1 \,+\,  a_s\, \left( 
-\frac{3}{2}\frac{1}{\epsilon} 
-\frac{3}{4} \right) 
\,+\, \mathcal{O}(a^2_s) \, .
\end{align}
We denote the quadratic Casimir in adjoint representation of SU($n_c$) gauge group by $C_A$, the number of active light quark flavor by $n_f$ and the leading coefficient of QCD $\beta$ function by $\beta_0 = 11C_A/3 -2n_f/3$.

To summarise, the renormalised effective Lagrangian with a non-anticommuting $\gamma_5$ for computing the $q\bar{q} \rightarrow ZH$ in HEFT can be written as
\begin{align}
\label{eq:Rlagrangian}
\mathcal{L}_{R} = \Big[\mathcal{L}_{c}+\mathcal{L}_{\mathrm{heff}}\Big]_{R} + 
\kappa~\mathrm{Z}_5^{h}(a_s)\, \mathrm{\mathbf{C}} \, \Big(\bar{q}_R(x) \,\left[\gamma^{\mu} \gamma_5\right]_{L} \, q_R(x)\Big) Z_{\mu}(x)\, H(x)\,,
\end{align}
where $\kappa_{ns} = c_t g_{A,q}$ and  $\kappa_{s} = c_t g_{A,b}$. The second term is the new amendment which does not arise while computing non-anomalous set of diagrams employing the anticommuting $\gamma_5$ scheme.

\section{Conclusions}
\label{sec:concl}
Through this article, we demonstrate two salient features of the $ZH$ amplitudes through quark annihilation. (1) The non-singlet (non-anomalous) part of the axial form factors of the $b\bar{b} \rightarrow ZH$ can be computed from a set of vector form factors, thus by completely bypassing the handling of $\gamma_5$.
(2) While computing the $q\bar{q} \rightarrow ZH$ amplitudes in HEFT by employing a non-anticommuting $\gamma_5$ in dimensional regularization, we need to introduce a four-point effective composite operator to the renormalized Lagrangian in order to fulfil the expected Ward identity.

\section*{Acknowledgements}
The author, together with Werner Bernreuther, Long Chen, and Micha\l{} Czakon thank the organisers of EPS-HEP2021.

\paragraph{Funding information}
The author received funding from the European
Research Council (ERC) under the European Union’s Horizon 2020 research and innovation programme \textit{High precision multi-jet dynamics at the LHC} (ERC Condsolidator grant agreement No 772009).


\setlength{\bibsep}{0pt}


\bibliography{skeleton.bib}
\bibliographystyle{utphysM}

\end{document}